# Evolution of the personal income distribution in the USA: High incomes

Ivan O. Kitov

## Abstract

The personal income distribution (PID) above the Pareto threshold is studied and modeled. A microeconomic model is proposed to simulate the PID and its evolution below and above the Pareto income threshold. The model balances processes of income production and dissipation for any person above 15 years of age. The model accurately predicts the observed dependence of the number of people reaching the Pareto threshold on work experience and the functional dependence of the relationship on the per capita real GDP growth for the period from 1994 to 2002. Predictions of the income distribution depending on age are given for past and future. In future, relatively less rich people are observed in the younger age groups and the peak of the relative number shifts to older ages with time. The effect of the power law distribution extending itself to very high incomes is speculated to be the cause of low performance of socialist countries.

## Introduction

A microeconomic model describing the personal income distribution (PID) in the USA was recently developed [1]. The basic approach draws from the field of geomechanics and reproduces some general features of the model for a solid with inhomogeneous inclusions [2]. The microeconomic model predicts the individual income evolution dependence on time when the person entered the economy, the work experience, the personal capability to earn money, and the effective size of the means used to earn money. When aggregated, the model predicts the PID and its evolution in time for the population above 15 years of age.

The principal assumption of the model is that there is no difference between money production and money earning. In contrast to the conventional economic theories, no one produces any profit for the rest of the society s/he belongs to. So, the amount of money produced by a person in the form of goods or services is exactly equal to the amount of money earned by the person. This condition also means that there is no money that is not produced by people, i.e. the total income or GDP exactly equals the sum of the personal incomes. In turn, the personal income distribution is fixed in relative terms at a characteristic time length of several years.

One of the essential features of the observed PID described by the model is the existence of two inherently different regimes of money earning. The first regime, referred to as "subcritical", corresponds to money income proportional to the (numerically defined) capacity of a person to earn money, i.e. to the product of the capability to earn money and the size of the earning means [1]. One can consider this capacity as the effective power of money production that the person achieves by applying her/his personal capability to earn money by some mechanism (leverage) for multiplying the personal power to earn money.

The second regime, a “supercritical” one, corresponds to the personal income range described by a Pareto or power law distribution. This regime starts at some non-zero income threshold, the Pareto threshold, and is supposedly the result of a self-organized-criticality (SOC) [3]. The personal income distribution here is governed by stochastic processes and any person in this income interval can reach any income according to a power law probability. No individual capability above the Pareto threshold is important for the final result, just the probability law. However, the Pareto threshold has to be reached first by passing through the subcritical branch. The Pareto distribution has several analogies in natural and social sciences.

Mechanisms leading to the power law distributions (scale free distributions of sizes - personal incomes in the case of economics) are studied in more detail in some natural sciences. The nature of such mechanisms resulting in the power law distribution in the field of economics is still a big challenge and is not considered here. One can assume, however, that the mechanisms work rapidly and there is no delay between the moment when some personal income reaches the threshold and the moment when the income jumps to its new position in the Pareto distribution. In seismology, for example, the final size or magnitude of an earthquake is usually reached several seconds after the rupture starts. Seismic magnitude represents the size of earthquakes, and this is also distributed according to a power law.

There are several problems related to the Pareto distribution that one can resolve in the framework of the developed model. One important problem is to determine the income threshold separating the subcritical and supercritical zones of income behaviour. There is a relatively wide transition zone between the two branches where the sub- and supercritical distributions practically coincide. The model accurately distinguishes the zones by matching various characteristics of the observed and predicted distributions.

The relative number of people having incomes in the supercritical zone depends on work experience. The relative number grows exponentially up to some critical age and then drops exponentially. This complex behaviour also explained by the model. Moreover, fine changes in the shape of the personal income distribution are predicted.

The model does not depend on any conventional economic theory or approach. It simulates some independent observations made in the USA during the last ten years. The observations are obviously carried out for some economic purposes and are based on some assumptions adapted from economics. For example, income is divided into personal and corporate income, despite the fact that the latter also belongs to some selected people on a personal basis. However, one can formally introduce a model linking these data sets that does not take into consideration any external meaning of the data. Even if the model has no deep economic roots, its merits have to be assessed by its predictive and resolution capabilities.

## 1. The personal income distribution in the USA. Observations

The personal income distribution is an economic parameter, which has been measured in the USA with increasing accuracy for more than fifty years [4]. Detailed data on personal income with age are available only from 1994, however [5]. The data are presented as the number of people falling within $2500 wide income intervals, starting at $0 and loss and ending with $100,000. For the years after 2000, the number of people with income in $50,000 wide intervals is published up to a maximum income of $250,000. This modification is obviously necessary because of an increasing number of people with incomes above $100,000, as discussed below. The data are also given in various age intervals. The official age of starting work in the USA is 15 years. The first age interval for the personal income statistics is 10-years wide and spans from 15 to 24 years of age. This interval corresponds to the work experience interval from 0 to 9 years. Further age intervals are 5-year wide and span the age ranges from 25 years to 74 years. There is an open-end interval above 75 years of age for all the individuals above this age. Complimentary estimates in 10-year wide intervals are also available. Hence, the data are represented as a 2-D table over age (work experience) and income with resolutions of 5 years (except the first interval) and $2,500 respectively.

Accuracy of observations is a key question for any modelling. If the measured data are not reliable, any relationship obtained from an observation/prediction matching procedure may be misleading. The U.S. Census Bureau provides some estimates of the PID accuracy in the form of a standard deviation corresponding to the observations [6]. These estimates, however, should be used with some precautions, as discussed in [6], taking into account the accuracy of the population estimates in various age groups. The population estimate accuracy is sometimes as low as 3% to 5% [7, 8].

The overall PIDs from 1994 to 2002 are shown in Figure 1. The original annual distributions of the absolute number, as given in the U.S. Census Bureau tables, are normalized to the total population for the corresponding calendar year in order to avoid potential side effects due to the population change, which may be as much as 1.5%. The obtained population density distributions are adjusted for the nominal per capita GDP growth according to a procedure formally equivalent to a standard adjustment for inflation. The normalized and adjusted curves demonstrate a high degree of similarity. This effect may be interpreted as an existence of a rigid structure for the PID. So, a portion of the total population having a given share of the total (nominal) income is constant in time. Thus any change in population distribution induced by any actual demographic processes and any change in economic growth (positive or negative) creates in the society forces that cause redistribution in the total income, which effectively returns the normalized and adjusted PID to the original one.

There are two clear parts of the PID distinguished in Figure 1: the initial quasi-exponential part and the part successfully interpolated by a power law. The observed value of the power law exponent obtained by a standard regression analysis is -3.97. This value practically coincides with the corresponding theoretical value of -4.0 (for the population density distribution).

Some PIDs for 1994 are presented in Figure 2 as a function of work experience. The curves are normalized to the total population in the corresponding income interval and illustrate the evolution of the PID obtained in the open high-end income intervals with an increasing low-end threshold. The first PID corresponds to all the possible incomes. This interval includes also those people who are reported in the original Census Bureau table as having no income or having losses. The second curve corresponds to the PID which includes all the people with income above $10K, and so on. The last curve represents the open-end PID for people with income above $100,000. All the incomes in the original tables are given in current US dollars.

The evolution of the curves is remarkable. The first curve has its peak in the first work experience interval. The counted number of people is assigned to the center of the work experience interval. The majority of people entering the economy have very low income of between $0 and $10K during the starting 10 years. With an increasing low-end threshold, the curves are gradually transforming into a bell-like or unimodal shape, with the peak moving towards the work experience interval between 30 years and 39 years. For the low-end threshold above $60K, all the curves practically coincide.

At first glance, the part of the PID described by the Pareto distribution seems to be the simplest one. Really, the observed behaviour of the PID at higher incomes obeys a simple law and should not change in time, at least theoretically. The total number of people with income in the zone controlled by the power law distribution develops with time as a linear function of the nominal per capita GDP and population growth. This theoretical conclusion is confirmed by the observations. Figure 3 depicts the number of people with income above $100,000 for the period from 1994 to 2002. The linear regression line shown in the Figure demonstrates that the theoretical assumption is correct.

Figure 4 presents the normalized number of people who reached the income threshold of $100K as a function of work experience for the years from 1994 to 2002. There is no significant change in the distribution, considering the accuracy of the counting [6]. This type of behaviour is not as easy to interpret as the above observation.

## 2. Microeconomic model

The principal assumption of the microeconomic model is that every person above fifteen years of age (the official starting age for work) has a capability to work or earn money using some means, which can be a job, bank interest, stocks, etc. An almost complete list of the means is available in the US Census Bureau technical documentation [4] as the sources of income are included in the survey list.

The income or the total amount of money a person earns per unit time is proportional to her/his capability to earn money or work, $\sigma$. The person is not isolated from the surrounding world and the work (money) s/he produces dissipates (leaks) through interaction with the outside world, decreasing the final income. Analogously to similar cases observed in the natural sciences [2], the rate of the dissipation is proportional to the attained income level and inversely proportional to the size of the means used to earn the money, $\Lambda$. One can write a simple balance equation for a person earning money:

$$dM(t)/dt = \sigma(t) - \alpha M(t)/\Lambda(t) \quad \textbf{(1)}$$

where $M(t)$ is money income denominated in dollars [$], $t$ is the work experience expressed in some appropriate time units [time unit], $\sigma(t)$ is the capability to earn money

[\$/(time unit)]; and $\alpha$ is the dissipation coefficient expressed in units [\$/(time unit)]. The size of the earning means, $\Lambda$, is also expressed in [\$]. The general solution of the above equation, if $\sigma(t)$ and $\Lambda(t)$ are considered to be slowly changing in time, is as follows (where $\sigma$ and $\Lambda$ are effectively constant):

$$M(t)=(\sigma/\alpha)\Lambda(1-exp(-\alpha t/\Lambda)) \quad \textbf{(2)}$$

In the modelling, we integrated equation (1) numerically in order to include the effects of the change with time in values of $\sigma(t)$ and $\Lambda(t)$. Equations (2) through (4) are derived and discussed in detail below to demonstrate some principal features of the model. The equations represent the solutions of equation (1) in the case where we neglect the observed change of $\sigma(t)$ and $\Lambda(t)$ in all the terms except the exponential one.

One can introduce the concept of a modified capability to earn money as a dimensionless variable $\Sigma(t)=\sigma(t)/\alpha$. The absolute value of the modified capability, $\Sigma(t)$, and size of the earning means evolves with time as the square root of the per capita real GDP: $\Sigma(t)= \Sigma(0)sqrt(GDP(t)/GDP(0))$ and $\Lambda(t)= \Lambda(0)sqrt(GDP(t)/GDP(0))$, where $GDP(0)$ and $GDP(t)$ are the per capita real GDP values at the start point of the modelling and at time $t$ respectively. Then the capacity of a "theoretical" person to earn money, defined as $\Sigma(t)\Lambda(t)$, evolves with time as the per capita real GDP. Effectively, equation (2) states that the evolution in time of personal income depends only on the personal capability to earn money, the means used to earn money and the economic growth.

The modified effective capability to earn money, $\Sigma(t)$, and the size of earning means, $\Lambda(t)$, obviously have nonzero minimum values for all the persons, $\Sigma_{min}(t)$ and $\Lambda_{min}(t)$ respectively. One can now introduce relative values of the defining parameters in the following way: $S(t)=\Sigma(t)/\Sigma_{min}(t)$ and $L(t)=\Lambda(t)/\Lambda_{min}(t)$. An assumption is made that the possible relative values of $S(0)$ and $L(0)$ can be represented as a sequence of integer numbers from 2 to 30, i.e. only 29 different integer values of the relative $S(0)$ and $L(0)$ are available: $S_1$=2,…, $S_{29}$=30; $L_2$=2,…, $L_{29}$=30. This discrete range results from the calibration process described in [1]. The largest possible relative value of $S_{max}=S_{29}$=30=$L_{max}=L_{29}$ is only 15 (=30/2) times larger than the smallest possible $S=S_1$ and $L=L_1$ (the model minimum values $\Lambda_{min}$ and $\Sigma_{min}$ are chosen to be two times smaller than the minimum observed values of $\Lambda_1$ and $\Sigma_1$). Because the absolute values of $\Lambda$, $\Sigma$, $\Lambda_{min}$, and $\Sigma_{min}$ evolve with time according to the same law, the relative values $L_i(t)$ and $S_i(t)$, $i$=1,…,29, do not change with time. This means that the distribution of the relative capability to earn money and the size of the earning means is fixed.

One can substitute the product of the relative values $S$ and $L$ and the time dependent minimum values $\Lambda_{min}$ and $\Sigma_{min}$ for $\Sigma(t)$ and $\Lambda(t)$ in (2) **.** We also normalize the equation to the maximum values $S_{max}$ and $L_{max}$**.** The normalized equation for income $M_{ij}(t)$ for a person with the capability, $S_i$**,** and the size of the earning means, $L_j$**,** is as follows**:**

$$M_{ij}(t)/(S_{max}L_{max})=(\Sigma_{min}\ \Lambda_{min})(S_i/S_{max})(L_j/L_{max})(1-exp(-(\alpha/\Lambda_{min}\ L_{max})t/(L_j/L_{max}))) \qquad \textbf{(3)}$$

or

$$M'_{ij}(t) = \Sigma_{min}(t)\Lambda_{min}(t)S'_iL'_j\{1-exp[-(1/\Lambda_{min})(\alpha' t/L'_j)]\} \qquad \textbf{(3')}$$

where $M'_{ij}(t)=M_{ij}(t)/(S_{max}L_{max})$**;** $S'_i=(S_i/S_{max})$**;** $L'_j=(L_j/L_{max})$**;** $\alpha'=\alpha/L_{max}$, $S_{max}$=30 and $L_{max}$=30. Below we omit the prime indices. The term $\Sigma_{min}(t)\Lambda_{min}(t)$ is equal to the total growth of per capita real GDP from the start point for the work experience $t$. One can consider this term as a coefficient defined for every single year of work experience because this is a predefined external variable. We can thus always measure the personal income in units $\Sigma_{min}(0)\Lambda_{min}(0)$**.** Then equation (3') becomes a dimensionless one and the coefficient changes from 1.0 as the per capita real GDP evolves relative to the start year.

Equation (3’) represents the income for a person with the defining parameters $S_i$ and $L_j$ at time $t$ relative to the maximum possible personal income. The maximum income is obtained by a person with $S_{29}$=1 and $L_{29}$=1 at the same time $t$. The term $1/\Lambda_{min}$ in the exponential term evolves in time inversely proportional to the square root of per capita real GDP. This is the defining term of the personal income evolution, which accounts for difference between the start years of work experience. The numerical value of the ratio $\alpha/\Lambda_{min}$ is obtained by calibration for the start year of the modelling. The calibration assumes that one can always consider $\Lambda_{min}(0)$=1 (and $\Sigma_{min}(0)$=1 as well) at the start point of the modelling and calibrate only the factor $\alpha$. In this case, the absolute value of $\alpha$ depends on the start year.

A probability for a person to get a means of relative size $L_j$ is constant over all values of the relative means size. The same is valid for $S_i$, i.e. all the people of age above 15 years are distributed equally among the 29 groups for the capability to earn money.

Thus, the relative capacity for a person to earn money is distributed over the population above 15 years of age as products of $S_i$ and $L_j$, where the $S_i$ and $L_j$ are independent: $S_iL_j$={2×2/900, 2×3/900, …, 2×30/900, 3×2/900, …, 3×30/900, …, 30×30/900}. There are only 841 (=29x29) values of the normalized capacity available between 4/900 and 900/900. Some of the cases appear to be degenerate (for example, 2x30=3x20=4x15= …= 30x2), but actually all of them define different time histories according to equation (3'), where $L_j$ is present in the exponential term. In the model, no individual future income trajectory is predefined, but it can only be chosen from the set of the 841 predefined individual futures for each single year of age.

It is illustrative to compare two cases with the same potential level of income but different $L_j$ values. Such cases are, for example, ($S_1$=2/30, $L_{29}$=30/30) and ($S_{29}$=30/30, $L_1$=2/30). Then the ratio $\alpha/\Lambda_{min}$ in (3') is equal to 0.08, $\Lambda_{min}(0)$=1 and $\Sigma_{min}(0)$=1. The first case represents the income evolution for an incapable person occupying the most profitable position. In the second case, a person with a prominent potential to earn money is forced to use the smallest earning means. The curves corresponding to the two cases are shown in Figure 5. Income for the first person grows very slowly compared to that for the second person. The persons reach approximately the same level of income after 40 years and about 3 years, respectively. The size of earning means, $L_j$, defines the time history of the personal income because of the presence in the exponential term. A characteristic time of income growth in (3') can be defined as $t^*=L_j\Lambda_{min}/\alpha$. The larger the characteristic time the longer time is needed to reach the same relative level of income. In turn, the capacity to earn money, $S_iL_j$, defines the maximum potential income level for the given person. This level is an asymptotic one and is never reached.

One can not determine the numerical value of the dissipation factor $\alpha$ though independent measurement. Instead, a standard calibration procedure is applied. By definition, the maximum relative value of $L_j$ ($L_{29}$) is equal to 1.0 at the start point of the studied period, $t_0$. The value of $\Lambda_{min}(0)$ is also assumed to be 1.0. Thus, one can vary the value of $\alpha$ in order to match the predicted and observed PIDs, and the best fit value of $\alpha$ is used for further predictions. The range of $\alpha/\Lambda_{min}$ from 0.09 to 0.04 approximately corresponds to that obtained in the modelling of the PID for the time period from 1960 to 2002 [1], i.e. $\alpha$=0.09 for $t_0$=1960. The numerical value of $\Lambda_{min}$ changes during this period from 1.0 to 1.49 according to the square root of per capita real GDP growth during the period. The total per capita GDP growth from 1960 to 2002 is 2.22.

One can find that eventually more and more time is necessary for a person with the maximum relative values $S_{29}$ and $L_{29}$ to reach the Pareto threshold of income. The initial Pareto threshold is found to be $M_{Pareto}(0)$=0.43 and it evolves in time as per capita real GDP: $M_{Pareto}(t)=M_{Pareto}(0)(GDP(t)/GDP(0))$ [1].

When a personal income reaches the Pareto threshold, it undergoes a transformation and obtains a new quality to reach any income with a probability described by the power law distribution. This approach is similar to that applied in the modern natural sciences involving self-organized criticality (SOC). Due to the exponential character of the income growth the number of people with incomes distributed according to the power law is very sensitive to the threshold value, but people with high enough $S_i$ and $L_j$ can eventually reach the threshold.

If the money earning capacity, $S_iL_j$, drops to zero at some critical time, $T_{cr}$, in a personal history [1], the solution of (1) is:

$$M_{ij}(t)=M_{ij}(Tcr)\exp(-\alpha\,(t-Tcr)/\Lambda_{min} L_j)=$$
$$=\{\Sigma_{min}(t)\Lambda_{min}(t)S_iL_j(1-\exp(-\alpha Tcr/\Lambda_{min} L_j))\}\exp(-\alpha(t-Tcr)/\Lambda_{min} L_j) \qquad \mathbf{(4)}$$

The first term is equal to the income value attained by the person at time $T_{cr}$, and the second term represents an exponential decay of the income for work experience above $T_{cr}$. The observed exponential roll-off for the mean income beyond $T_{cr}$ corresponds to zero work applied to earn money in the model [1]. People do not exercise any effort to produce income starting from some predefined point in time, $T_{cr}$, and enjoy exponential decay of their income.

There is a principal feature of the real PID, which is not described by the model so far, but has an inherent relation to the studied problem. The real income distribution spans the range from $0 to several hundred million dollars, and the theoretical distribution extends only from $0 to about $100,000, i.e. the income interval used in [1] to match the observed and predicted distributions. The power law distribution starting from the Pareto threshold income (from $40,000 to $60,000 during last fifteen years) describes incomes of about ten per cent of the population. The theoretical threshold of 0.43 was introduced above, partly, in order to match this relative number of people distributed by the Pareto law. The model provides an excellent agreement between the real and theoretical distributions below the Pareto threshold. Above the threshold, the theoretical and real

distributions diverge. Figure 6 presents the theoretical and observed dependence on income of the cumulative number of people with incomes below a given income value for the year 1999. The curves start at the point (0,0), i.e. no people without income, and practically coincide up to the income value of $54K. This value is the determined real absolute value of the Pareto threshold for 1999, which corresponds to the dimensionless Pareto threshold value of 0.951 in 1999 and 0.430 at the start point of the modelling in 1960.

Above the absolute value of the Pareto threshold, the theoretical distribution drops with an increasing rate to zero at about $100,000. This limit corresponds to an absence of the theoretical capacity to earn money, $S_iL_j$, above 1.0. The dimensionless units can be converted into actual year 2000 dollars by multiplying using a factor of $120,000, i.e. one dimensionless unit costs $120,000. The observed distribution decays above the Pareto threshold inversely proportional to income in the fourth power. Hence, the real and theoretical absolute income intervals are different above the Pareto threshold and retain the same portion of the total population (~10%). Thus, the total amount of money earned by people in the Pareto distribution income zone, i.e. sum of all the personal incomes, differs in the real and theoretical cases. Here one can introduce a concept distinguishing below-threshold (subcritical) and above-threshold (supercritical) behaviour of the income earners. In the subcritical zone, the income earned by a person is proportional to her/his efforts or capacity $S_iL_j$. In the super-critical zone, a person can earn any amount of money between the Pareto threshold and the highest possible income. A probability to get a given income drops with income according to the Pareto law.

The total amount of money earned in the supercritical zone is about 1.35 times larger than the amount that would be earned if incomes were distributed according to the theoretical curve, in which every income is proportional to the capacity. Figure 7 illustrates the concept. The two curves in the Figure correspond to the theoretical and observed total income received by the people with the incomes below a given value, i.e. the sum of all the personal incomes from a given value $M$ to zero income. The theoretical curve is not corrected for a 35% increase for each personal income above the Pareto threshold.

This multiplication factor is very sensitive to the definition of the Pareto threshold. In order to match the theoretical and observed total amount of the money earned in the supercritical zone one has to multiply every theoretical personal income in the zone by a factor of 1.35. This is the last step in equalizing the theoretical and the observed number of people and incomes in both zones: sub- and supercritical. It seems also reasonable to assume that the observed difference in distributions in the zones is reflected by some basic difference in the capability to earn money.

So, the model is finalized. An individual income grows in time according to relationship (3') until some critical age $T_{cr}$. Above $T_{cr}$, an exponential decrease according to (4) is observed. In modelling of average personal income, when an income is above the Pareto threshold it gains 35% of its theoretical value [1] in order to fit the overall income above the Pareto threshold. It is obvious that if a personal income has not reached the Pareto threshold before $T_{cr}$, it never reaches the threshold because it starts to decay after this length of work experience. Lets a personal income is above the Pareto threshold at work experience below $T_{cr}$. It drops exponentially with work experience above the $T_{cr}$. If it reaches the Pareto threshold is loses its extra 35% value.

All the population above 15 years of age is divided into 29 groups according to the capability to earn money. Any new generation has the same distribution of $L_j$ and $S_i$ as the previous one, but different start values of $\Lambda_{min}$ and $\Sigma_{min}$ which evolve with the real GDP per capita. The actual PID depends on the single year of age population distribution. The population age structure is an external parameter evolving according to its own rules. The critical work experience, $T_{cr}$, also grows proportionally to the square root of per capita real GDP [1]. Based on independent measurements of the population distribution and GDP one can model the evolution of the PID below and above the Pareto threshold.

## 3. Modelling high incomes

The best-fit distributions in the high income zone in the USA are obtained from a model with the following defining parameters [1]: the start year is 1960, $T_{cr}(1960)$=26.5 years, $\alpha$=0.087, $\Lambda_{min}(1960)$=1.0 and $\Sigma_{min}(1960)$=1.0, and the initial value of the Pareto threshold $M_{Pareto}(1960)$=0.43. The model fits various observations including the overall PID and the PIDs in various age groups, their evolution in time for the period from 1994 to 2002, the average income dependence on work experience and its evolution in time for the period from 1967 to 2002, and the evolution of $T_{cr}$ in time from 1967 to 2002. The model also predicts the behaviour of the PID in the high-income zone, above the Pareto threshold.

The number of people reaching the Pareto threshold depends on work experience. In the youngest age group, one can not expect a large number of people with high income. A comparison of the modelled and observed PID for the youngest group, as shown in Figure 8, confirms this assumption. In the age groups well above the critical time, $T_{cr}$, there are also less and less rich people with age, in absolute and relative terms. The peak of the Pareto distribution dependence on work experience is near $T_{cr}$, which is of about forty years currently.

Figure 8 displays the predicted and observed number of people with income above the Pareto threshold as a function of work experience in 1994 and 2002. The Pareto threshold is evolving in time as the per capita real GDP. The initial normalized Pareto threshold is 0.43 in 1960. It reaches a value of 0.829 in 1994 and 0.953 in 2002. Relationships (3') and (4) were used to predict the personal income for every person above 15 year of age in 1994 and 2002. The population distributions were obtained from the U.S. Census Bureau web-site [9]. The model curve is slightly lower than the observed one in 1994, but the curves for 2002 are in excellent agreement. However, the limited accuracy and resolution of the income and population measurements is noteworthy. The discrepancy between the theoretical and observed curves might be induced in part by the inaccuracy of the measurements. The income distribution resolution of \$2500 also adds to the discrepancy. One can only use the personal income distribution data with a step of \$2500 when the total change of the nominal Pareto threshold from 1994 to 2002 was of \$14000 – from \$43.5K to \$57.6K respectively. For example, we used the income interval above \$42.5K to present the actual distribution for what obviously overestimated the total number of people in any work experience interval, because all the people with incomes between \$42.5K and \$43.5K were counted in. Resolution for the theoretical income distributions is \$1K and 1 year of age.

The most important result of the comparison is that the shapes of the observed curves are very well predicted. This demonstrates the adequacy of the model in describing the underlying physical and social processes governing the features of the Pareto distribution.

Figure 9 presents the relative number of people who reached an income of \$100K (current dollars) as a function of work experience for the years 1994 and 2001. The predicted curves are consistent with the observed ones in shape and level. There is a slight difference in the initial parts of the theoretical and actual distributions. More people are measured in the very first work experience group in comparison with the predicted values. Despite a minor influence of the observation on the overall distribution in terms of the total income, one can argue that this difference is due to a wrong dissipation factor used in the model. This factor defines the time constant and the rate of income growth. The initial part of the distribution is most sensitive to the factor value. We have no convincing explanation for the discrepancy but just mention that the real start point for some people is well below 15 years of age and the accuracy of measurements at higher incomes for the youngest group is definitely low. Also, the income data resolution is very low: the first interval is 10 years wide. A data set with a finer resolution could help to reveal the reason for the discrepancy.

There are no observations published by the U.S. Census Bureau for the personal income distribution in the high income zone before 1994. Thus, one can not compare the observed and predicted distributions before this year, but one can reveal some important features of the theoretical distributions. Figure 10 depicts theoretical curves of the population density income distribution above the Pareto threshold as a function of work experience for some years between 1980 and 2002. The curves are normalized to the total population in the Pareto income zone for the corresponding year and present a clear picture of the evolution during the modelled period. In the beginning of the 1980s, when the effective dissipation factor, $\alpha/\Lambda_{min}$ was as large as 0.08, the time needed to reach the Pareto threshold of about 0.64 was lower and people relatively easily attained this level in the first 10 years of work. This corresponds to an almost linear growth in the number of people reaching the threshold in the first years of work. With increasing $\Lambda_{min}$, the effective dissipation factor was decreasing and the time needed to reach the threshold was eventually growing. The start segment of the curves became less steep and the principal growth migrated from the first decade to the second decade of work experience. This stage of the evolution has been observed during the last 10 years. It would be very interesting to compare the initial part of the theoretical distribution in the late 1970s and early 1980s with any reliable observations, if available.

## 4. Discussion and conclusions

The model predicts the exact number of people reaching the Pareto threshold depending on work experience. The distribution evolves with time in a manner also predicted by the model. Thus the model can be also used to predict future development of the Pareto distribution if the evolution of the population structure and per capita GDP are available. Figure 11 presents such a prediction for the years from 2002 to 2023 based on the population projection published by the U.S. Census Bureau [10] and per capita GDP growth trend of 1.6%. This prediction is possible because there is no other random or deterministic process which leads to the power law distribution except the process of the personal income to reach the Pareto income threshold as described by the model. (Specific realization of self-organized criticality for the personal incomes results in the Pareto or power law distribution.) In fact, if such a process does exist and adds (subtracts) sufficient number people to the observed one with incomes above the Pareto threshold, on could not expect that the predicted value would match the observed one. Hence the derived functional dependence on work experience would not match the observed data.

The model explicitly states that if all the positions in the Pareto distribution are occupied there is no opportunity to create a new one and occupy it. They are all enumerated and limited. People can swap their positions, however. If a person has low capability to earn money, there is no way to get rich. If s/he has a high capability but a small-sized earning means, s/he is potentially capable to change the means to a bigger one and reach the Pareto threshold. When one's personal income reaches about $57,000 in 2005 it is a good start to reach any income from the observed distribution with the probability inversely proportional to the income cubed as described by the Pareto distribution. Because the personal income distribution is predefined and rigid, the majority of the people can never reach the Pareto threshold and have the possibility to get rich. The majority, about 90%, is always beyond the Pareto threshold and gets income exactly proportional to the personal capability to earn money.

This is a good place to briefly discuss capitalism and socialism as economic systems. The latter economic system is based on an assumption that personal income is proportional to the time necessary to produce some goods or service and to some varying in the people capability to produce. This is the principle of socialism as an economic system - to obtain exactly proportional to the product produced. The price of the product is determined by some economic authority according to some rules aimed to balance inputs of time and productivity of the population. As we have seen above, this assumption works excellently for the overwhelming majority. Ninety per cent of the population above 15 years of age in the USA receives income exactly proportional to their capability to produce income, as described by the microeconomic model. When extended to the whole population, this rule limits personal incomes of the ten per cent of the population having incomes above the Pareto threshold to theoretical values determined by the model or lower. Capitalism, however, has some extra feature – the ten per cent of the population have personal income not proportional to the capability, but described by a power law spanning to several million dollars. They actually produce some extra income exceeding the theoretical value by 35%. The actual total income produced by these ten per cent is 45% of the actual overall total income, *R*, instead of the theoretical value of 34% of the predicted total income, *P*. Because the total income in the subcritical zone is equal in the real and theoretical cases $P(1-0.34)=R(1-0.45)$, the ratio $R/P=1.2$ or the actual total income exceeds the theoretical one by 20%. Hence, capitalism has an advantage of the personal income distributed by the Pareto law, which effectively increases the overall income or real GDP at least by 20% compared to that of socialism. In the long run, the extra total income provides progressively increasing additional GDP. Hence, the developed capitalist countries grow with a higher rate than socialist countries.

It is worth noting that the presented model is a "first principles" model. It does not approximate or interpolate the observed data but predicts functional dependence of the defining parameters. An essential feature of the model is its simplicity: there is only one first order ordinary differential equation defining any individual income trajectory. Also, the model has deep roots in natural sciences that suggest that economic activity is just a natural process governed by laws inherently following physical laws. Economics often considers human behaviour as unpredictable and even stochastic. A good analogue of such a system in physics is an ensemble of gas particles in a box. Nobody can predict a trajectory for a given particle. One can predict, however, the most probable number of particles in any given energy or velocity range and such macro parameters as temperature and pressure.

## Acknowledgments

The author is grateful to Dr. Wayne Richardson for his constant interest in this study, help and assistance, and fruitful discussions. The manuscript was greatly improved by his critical review.

## References

1. Kitov, I.O. “A model for microeconomic and macroeconomic development.” Working Papers 05, ECINEQ, Society for the Study of Economic Inequality.
2. Rodionov, V.N., V.M.Tsvetkov, I.A.Sizov. Principles of Geomechanics. Moscow. Nedra, 1982, pp.272. (in Russian)
3. Lise S. and M. Paczuski. A Nonconservative “Earthquake Model of Self-Organized Criticality on a Random Graph”, cond-mat/0204491, v1, 23 Apr 2002, pp. 1-4.
4. U.S. Census Bureau; "Current Population Survey. Design and Methodology". Issued March 2002 TP63RV. http://www.census.gov/prod/2002pubs/tp63rv.pdf
5. U.S. Census Bureau; "Detailed Income Tabulations from the CPS". Last Revised: August 26 2004; http://www.census.gov/hhes/income/dinctabs.html
6. U.S. Census Bureau, "Source and Accuracy of Estimates for Income in the United States: 2002" , http://www.census.gov/hhes/income/income02.sa.pdf
7. Kirsten K. West and J. Gregory Robinson, "What Do We Know About The Undercount of Children?", Population Division, U. S. Census Bureau Washington, DC 20233-8800 August 1999 Population Division Working Paper No. 39
8. U.S. Census Bureau, **"**Methodology: National Intercensal Population Estimates**". L**ast revised: August 20, 2004 at 07:19:10 AM. http://www.census.gov/popest/archives/methodology/intercensal_nat_meth.html
9. U.S. Census Bureau, U.S., “Population Estimates, Population Division” Maintained By: Information & Research Services Internet Staff (Population Division), http://www.census.gov/popest/estimates.php
10. U.S. Census Bureau, U.S. “Interim Projections by Age, Sex, Race, and Hispanic Origin.” Population Division, Population Projections Branch. Maintained By: Information & Research Services Internet Staff (Population Division) Created: March 18, 2004, Last Revised: August 26, 2004 at 02:48:52 PM http://www.census.gov/population/www/projections/popproj.html

## Figures

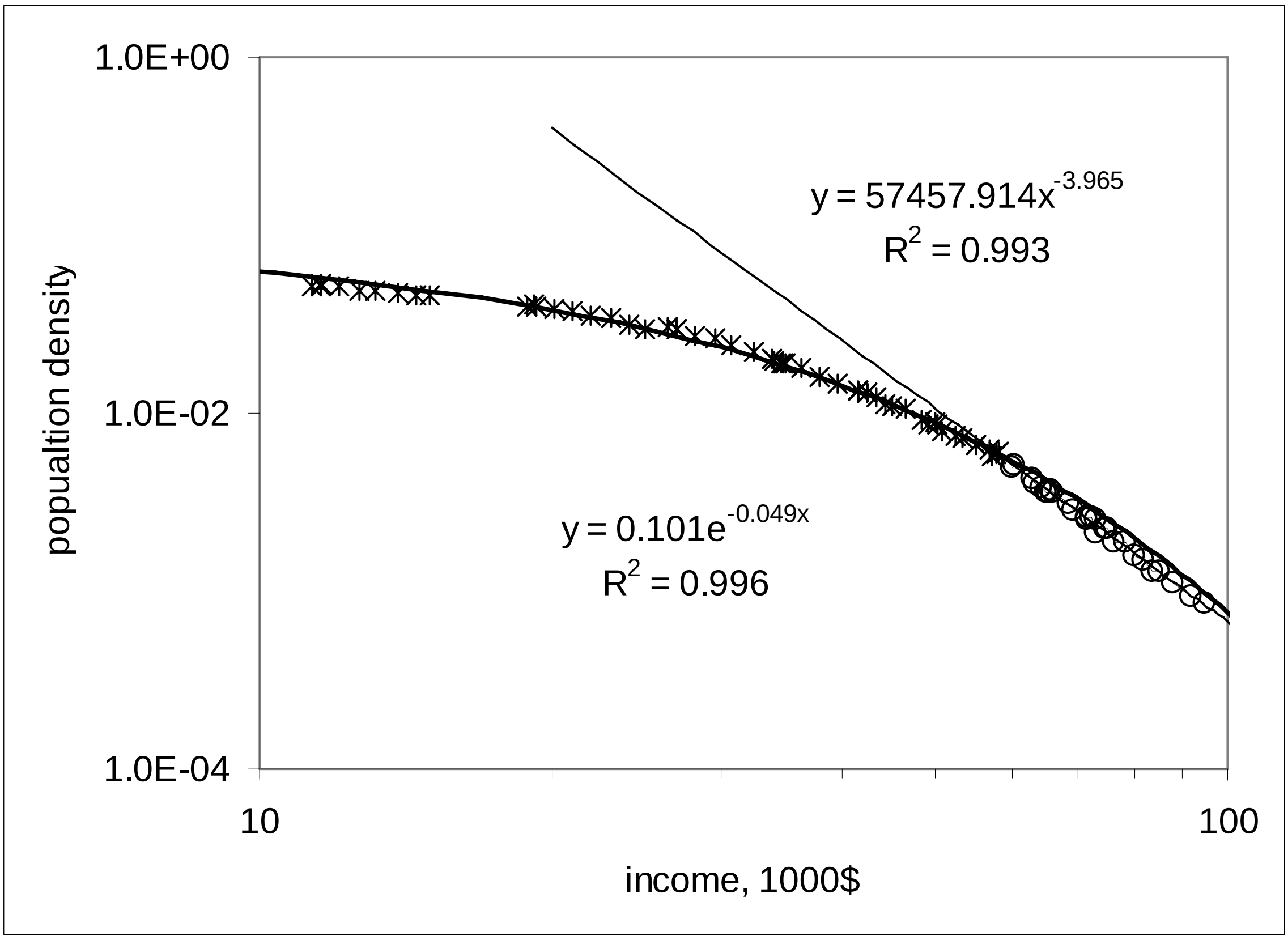

Fig. 1. Personal income density distributions below and above the Pareto threshold for years 1994 to 2002. The distributions are adjusted for the nominal per capita GDP growth. A power law regression demonstrates that the adjusted distributions practically coincide. The power law exponent -3.97 is almost equal to the theoretical value of -4.

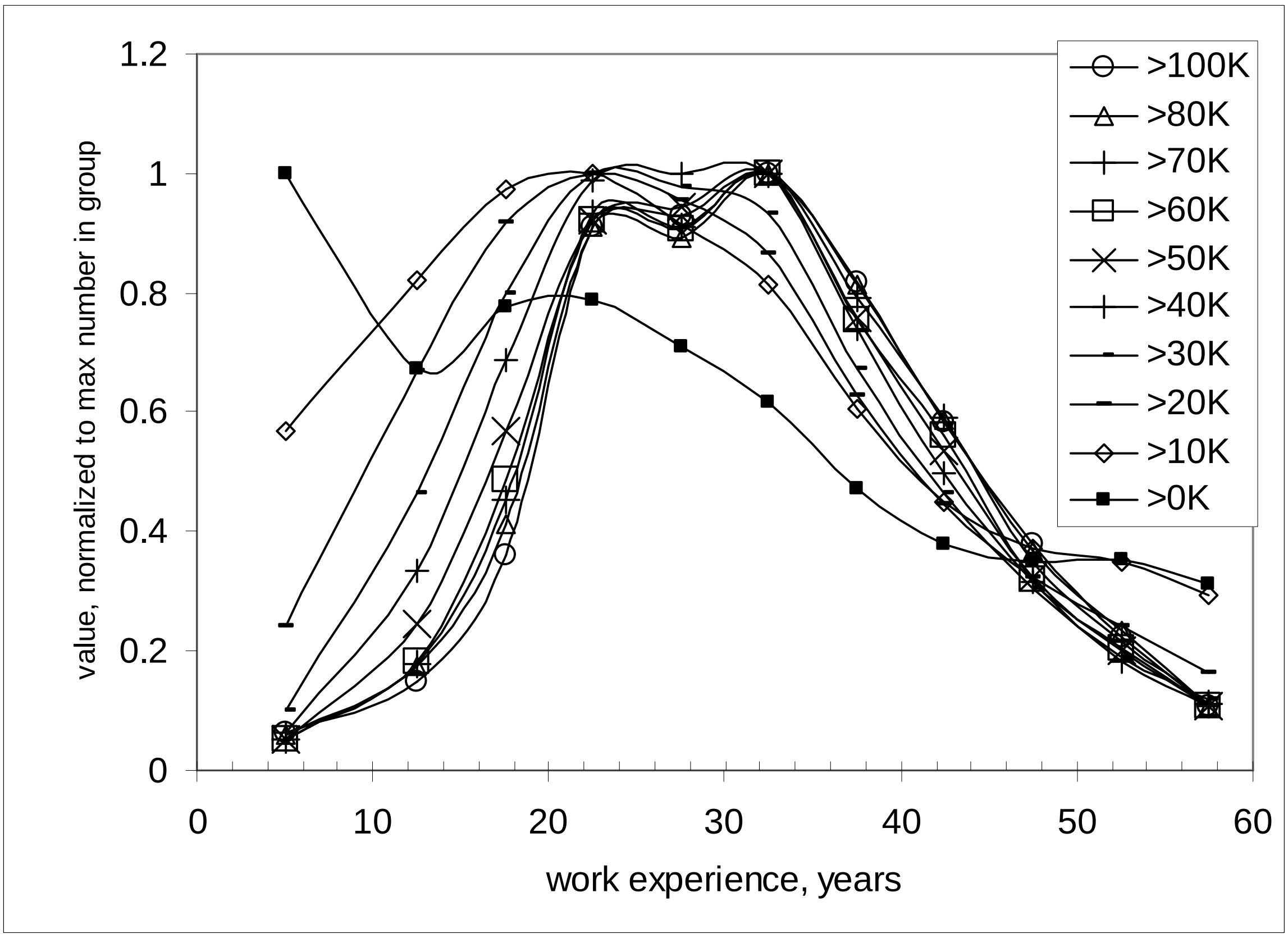

Fig.2. Evolution of the normalized personal income distribution in 1994 for incomes above a given threshold: >$0K (all personal incomes), >$10K, …, >$100K. The distributions for income above $60K are very similar.

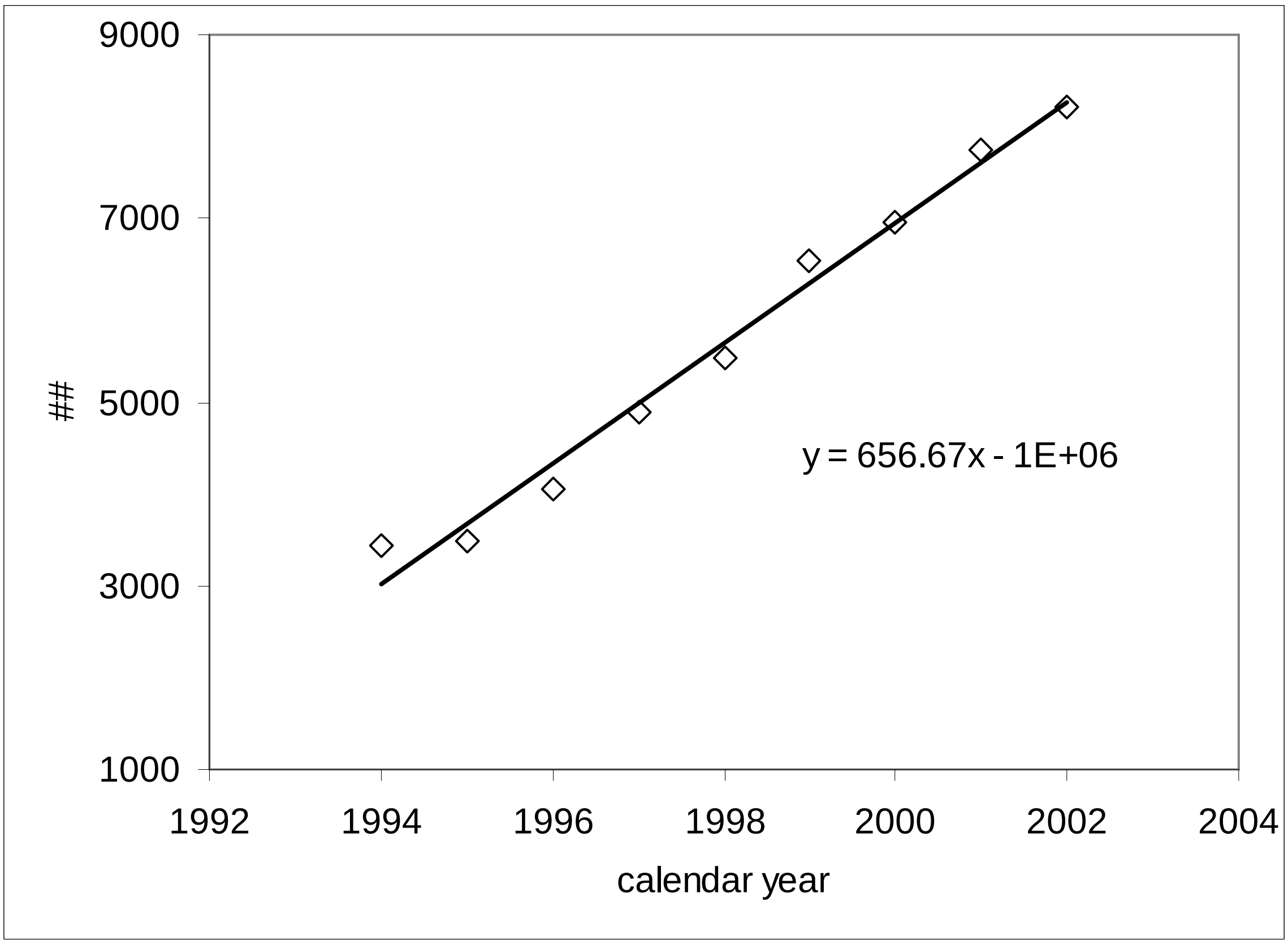

Fig. 3. The evolution of the number of people with income above $100K. A linear increase with nominal GDP growth is expected.

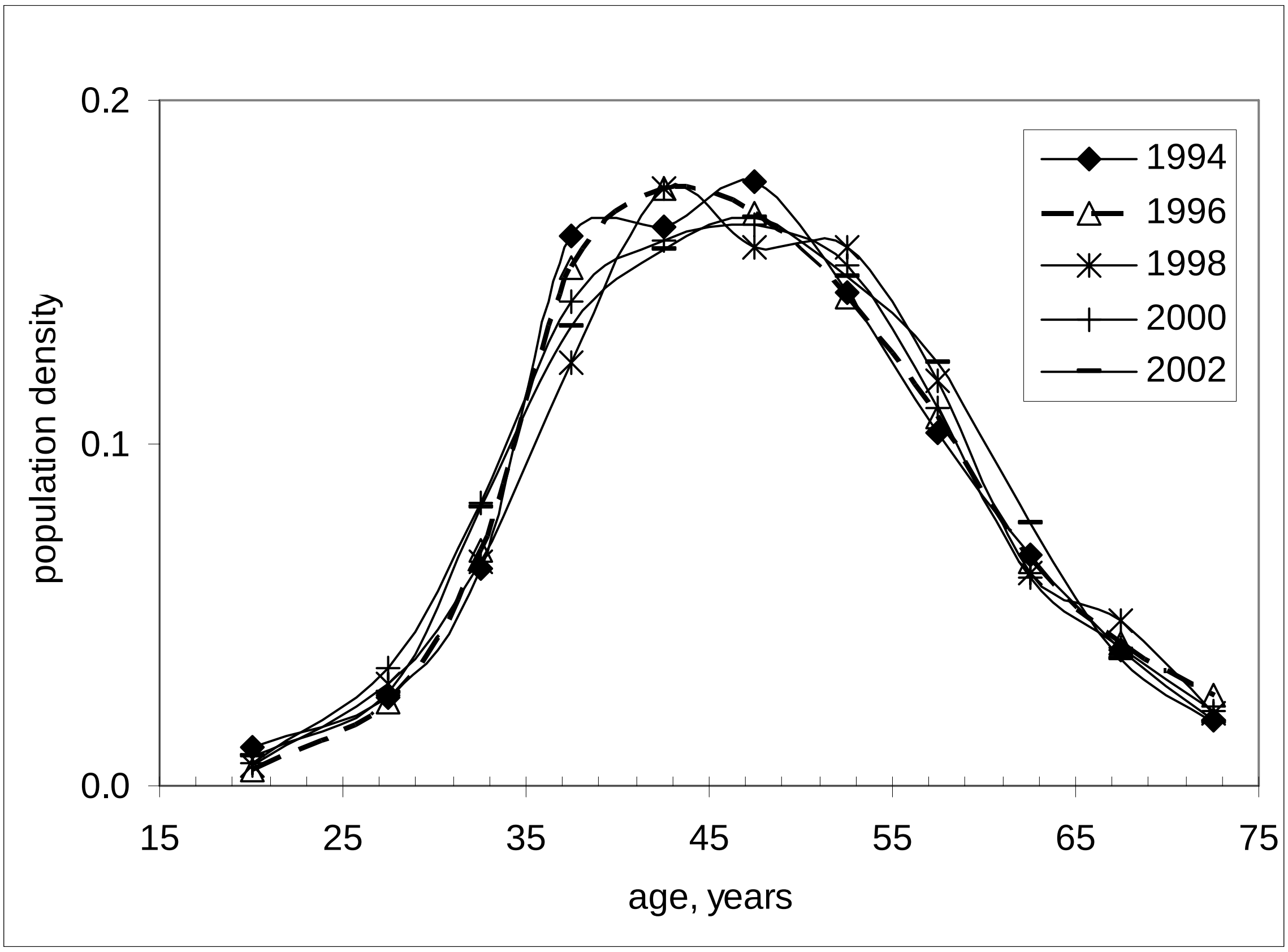

Fig. 4. Normalized personal income distribution for incomes above $100K as a function of work experience. The curves are shown for even years from 1994 to 2002.

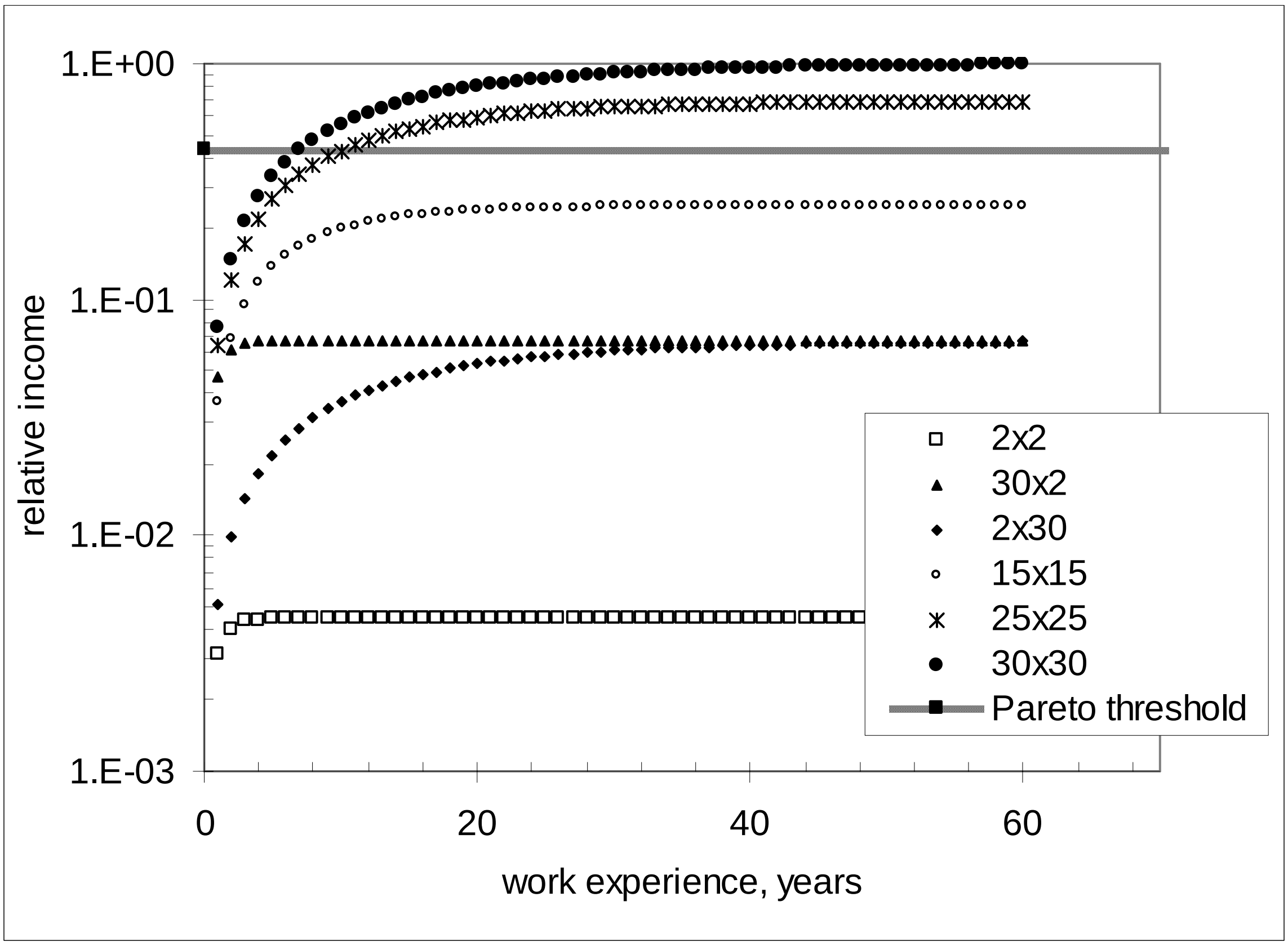

Fig. 5. Some trajectories of the theoretical income growth depending on the size of earning means, L, and the capability to earn money, S. Only the persons with high L and S can reach the Pareto threshold.

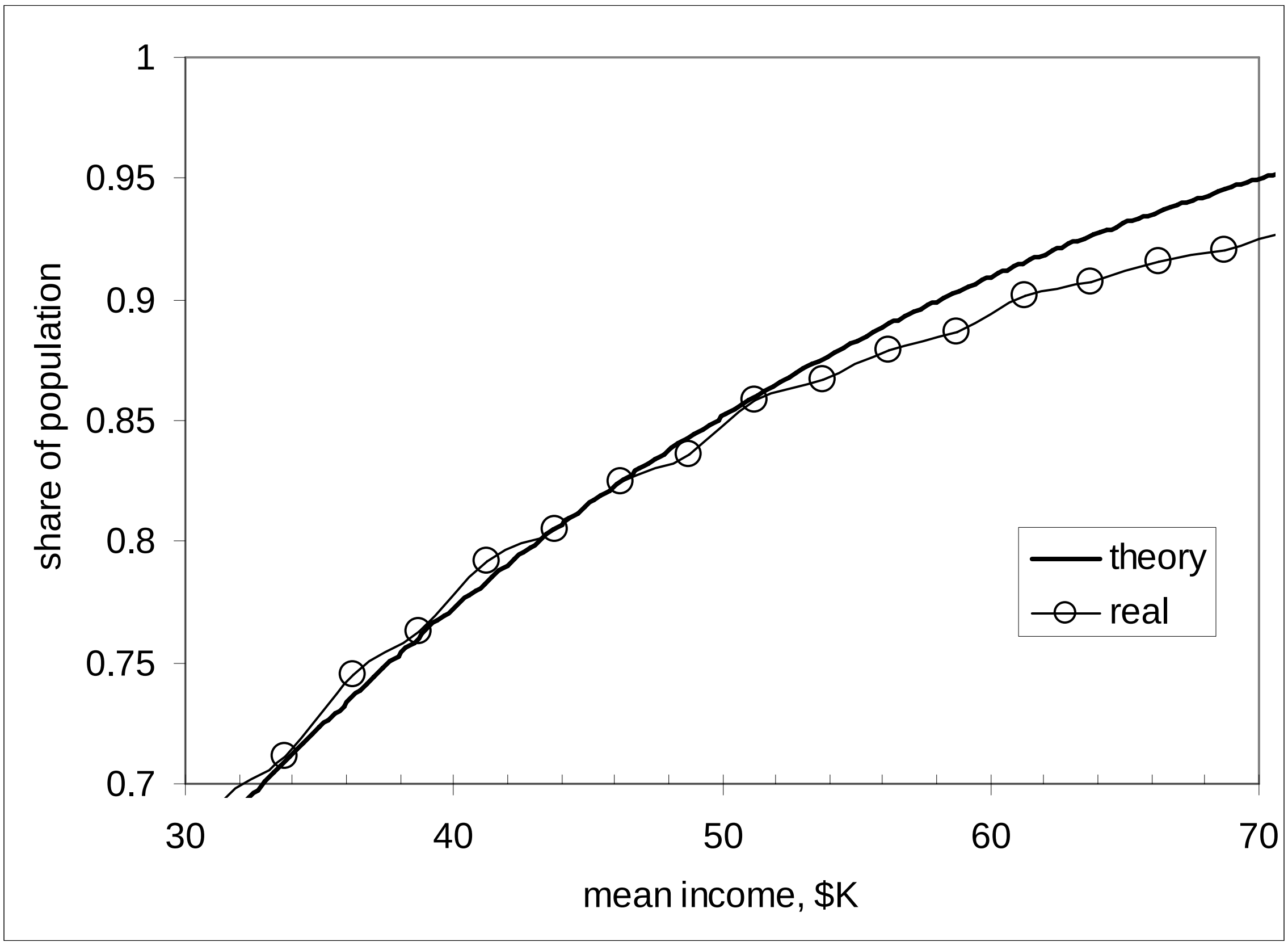

Fig. 6. Comparison of the observed and predicted personal income distributions for the year 1999 – a portion of population with income below a given value. The curves diverge at income of $54K – the Pareto distribution threshold.

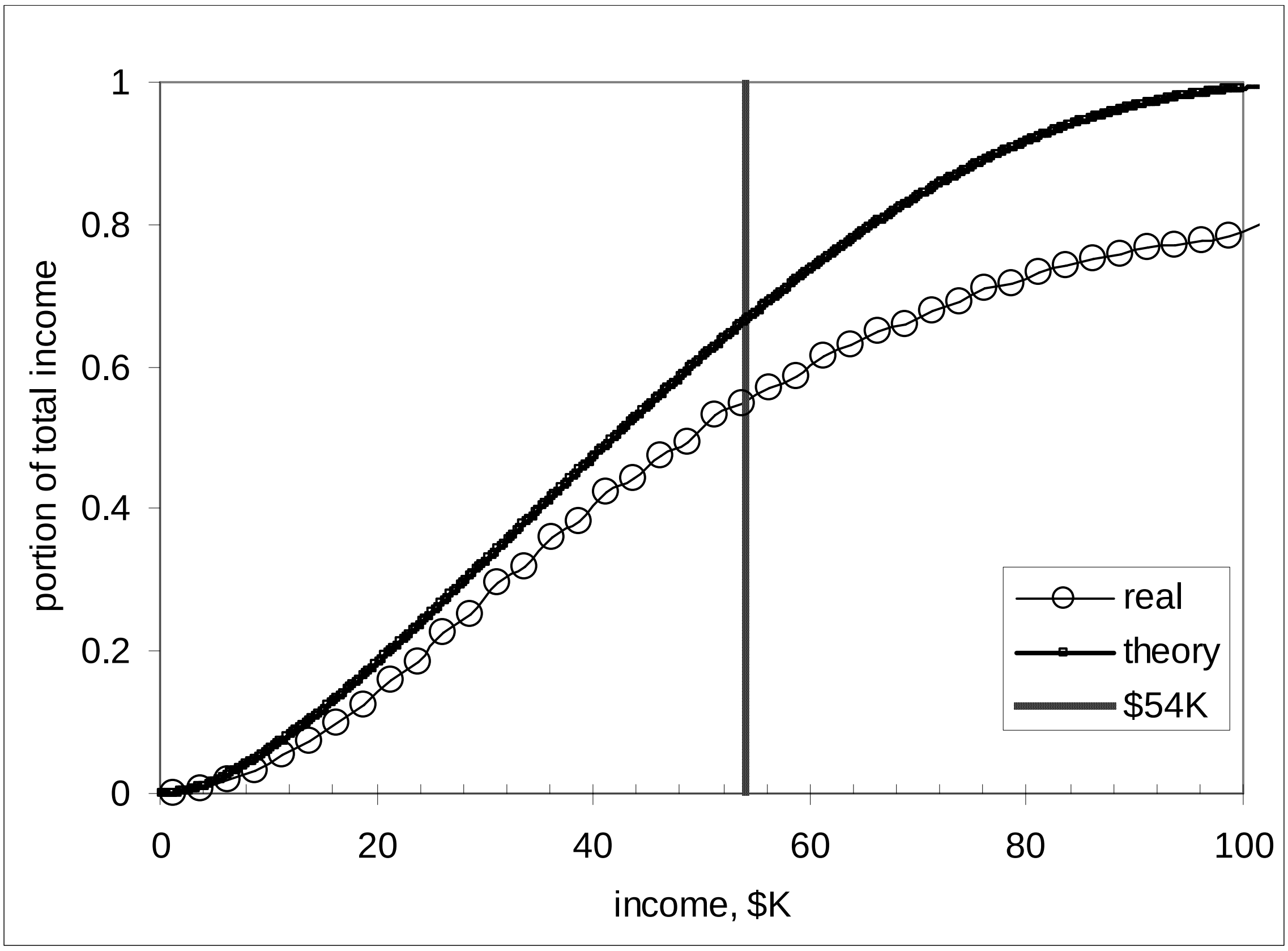

Fig. 7. Comparison of the observed and predicted cumulative personal income distributions for 1999 – a portion of total income received by population with income below a given value. The ratio of the observed cumulative income of the population above the Pareto threshold (0.450 – intercept of the vertical line and the solid curve) and the corresponding theoretical value (0.333 – intercept of the vertical line) and is equal to 1.35. This value is considered as an effective increase of the average capacity to earn money for people above the Pareto threshold.

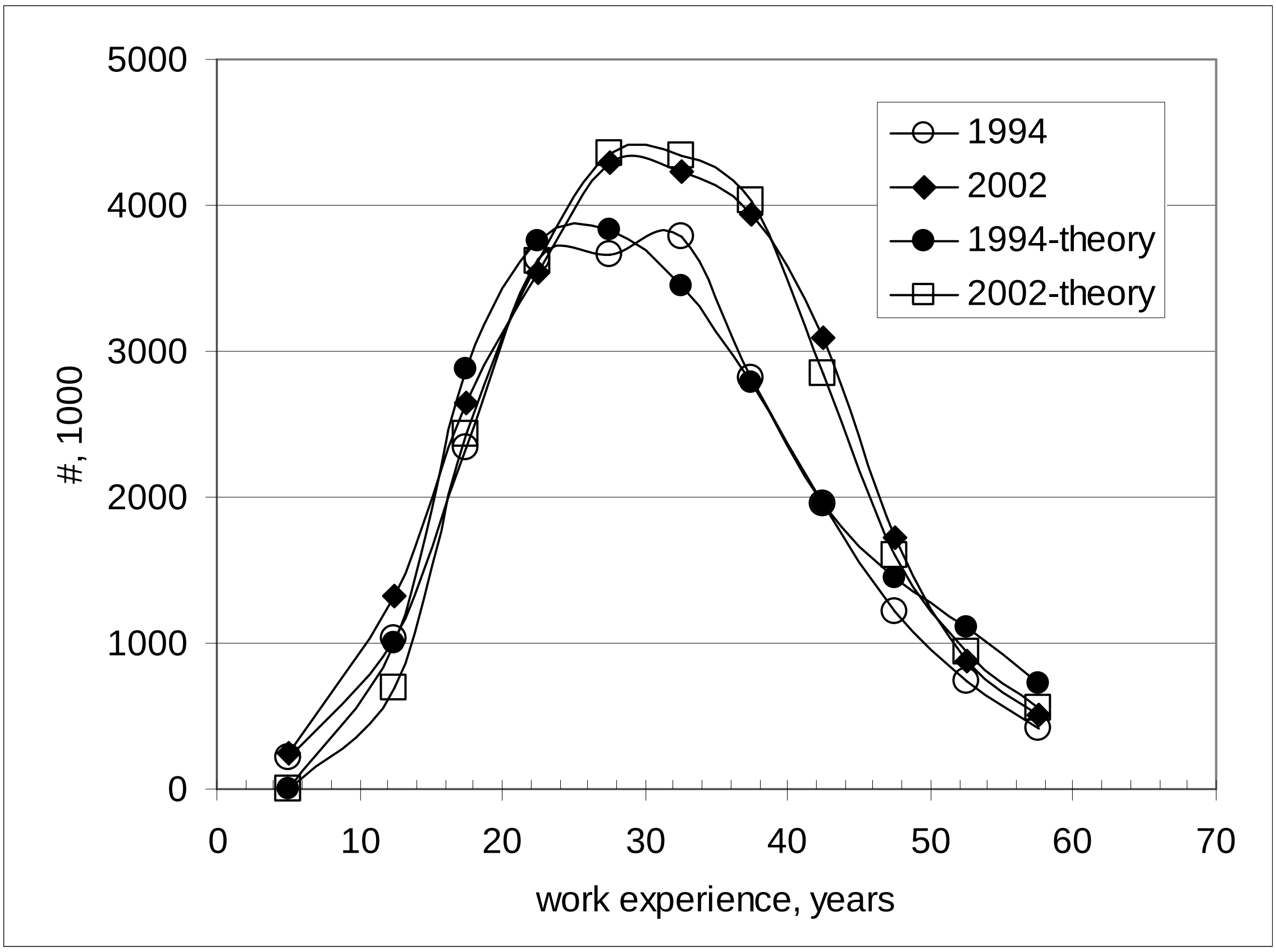

Fig. 8. Comparison of the observed and predicted dependence of the number of people with income above the Pareto threshold on work experience.

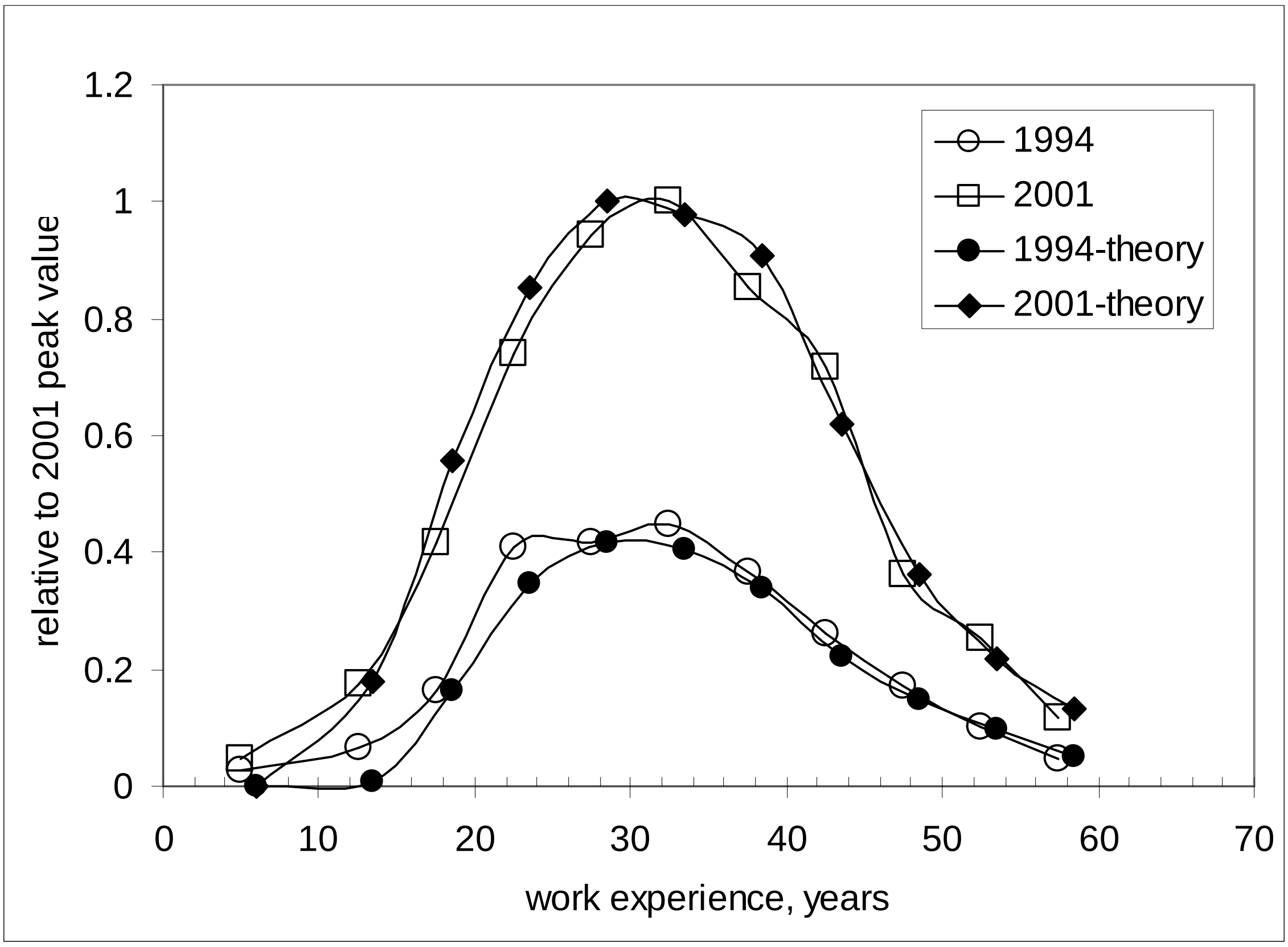

Fig. 9. Comparison of the observed and predicted dependence of the number of people with income above $100,000 (current dollars) on work experience. The curves are normalized to the peak value in 2001.

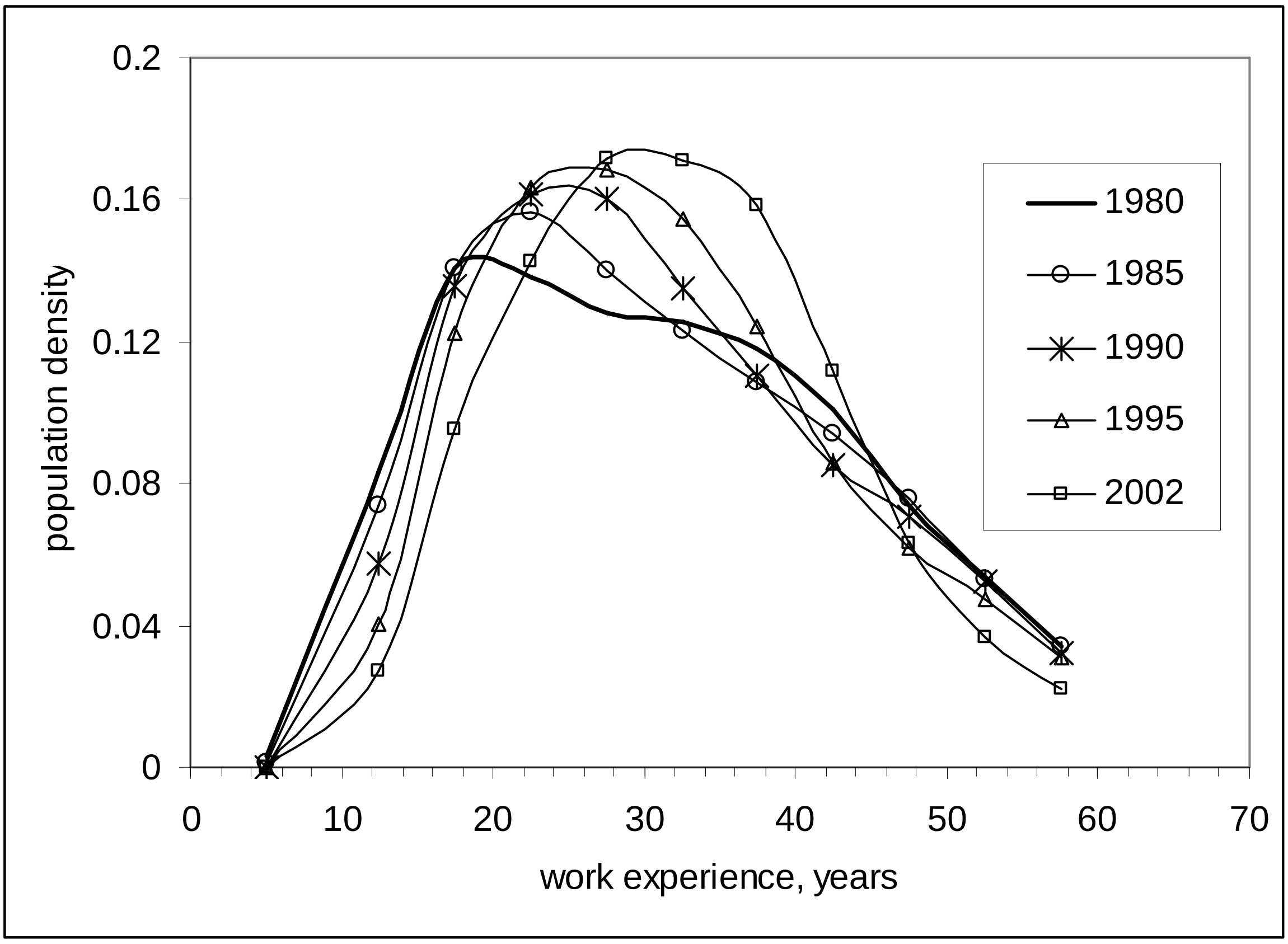

Fig. 10. Evolution of the normalized population density distribution for people with income above the Pareto threshold for calendar years 1980 through 2002. Note the increase of time needed to reach the Pareto threshold from 1980 to 2002. This is the result of the decrease in the effective dissipation factor, $\alpha$/L, with increasing size of earning means, L.

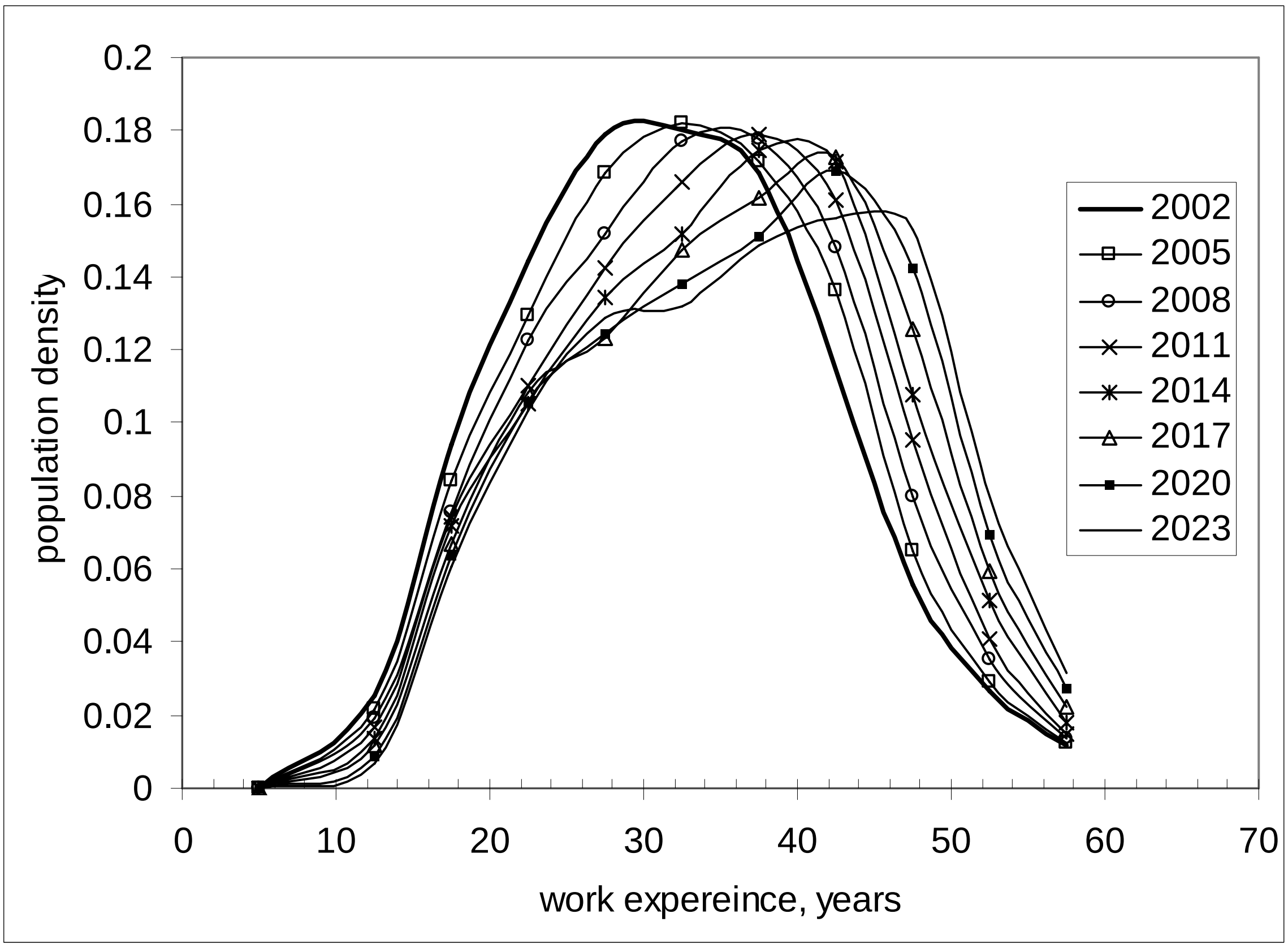

Fig. 11. Evolution of the normalized population density distribution for people with income above the Pareto threshold for calendar years 2002 through 2023. Per capita real GDP growth is set to be 0.016. Population projections obtained from U.S. Census Bureau [10] are used.